\documentclass[aps,pre,twocolumn,showpacs,10pt]{revtex4-1}
\usepackage{amsmath}
\usepackage{amsfonts}
\usepackage{amssymb}
\usepackage{epsfig}
\usepackage{graphicx}

\begin{document}
\title{Gelling by Heating}

\author{S{\'andalo} Rold{\'a}n-Vargas$^1$, 
Frank Smallenburg$^1$, 
Walter Kob$^2$, 
Francesco Sciortino$^1$
}
\affiliation{ $^1$Department of Physics, Sapienza, Universit\`{a}
di Roma, Piazzale Aldo Moro 2, I-00185, Roma, Italy,\\ $^2$Laboratoire
Charles Coulomb, UMR 5221, CNRS and Universit\'e Montpellier 2,
Montpellier, France}

\begin{abstract}
We introduce a simple model, a binary mixture of patchy particles,
which has been designed to form a gel upon heating. Due to the specific
nature of the particle interactions, notably the number and geometry
of the patches as well as their interaction energies, the system is a
fluid both at high and at low temperatures $T$, whereas at intermediate
$T$s the system forms a solid-like disordered open network structure,
i.e. a gel.  Using molecular dynamics we investigate the static and
dynamic properties of this system.

\end{abstract}

\pacs{
83.80.Kn, 
47.57.-s, 
64.70.qd, 
61.20.Ja} 

\maketitle

Some of the most versatile and efficient strategies for designing
new materials with unconventional behavior are  based on the idea
of {\it competitive interaction}. While in biology the term {\it
competitive interaction} indicates the result of rivalry between two
or more species competing for resources, in physics it stands
for the presence of several interaction mechanisms that can stabilize
competing local structures, leading to novel and highly interesting
features of the system.  Recent examples for this mechanism in soft
matter systems include, among others, the competition between short-range attraction
and long-range repulsion in charged colloids giving rise to cluster
phases~\cite{sear,kegelcluster,mossa}, the competition between chaining
and branching in  patchy colloids~\cite{safran,russoj} where a specific
design of the inter-patch interactions results in a phase diagram in
which the density of the coexisting liquid approaches the density
of the gas~\cite{russoj}, and the design of DNA-coated colloids
with two different DNA sequences for the purpose of establishing a
competition between intra and inter-particle interactions, favoring
crystal formation~\cite{frenkelnatmat}. Often the very nature of these
competing mechanisms promotes the emergence of a structure controlled
by energy (stable at low $T$) which competes with a structure stabilized
by entropy at intermediate $T$.

Recent progress in the synthesis of colloids have led to a new generation
of particles with highly directional and selective interactions,
providing valence to colloids~\cite{newsandviews}. The  ability to
tune the interactions between nano- and meso-sized particles almost at
will~\cite{glad1,dnapatchy1}, and to design the geometric properties of the
patches\cite{flavioNC} and/or their functionalization,  offers today the
possibility to exploit the idea of competitive interactions to modulate
material properties with  external control parameters.  In this Letter we
develop one such possibility: The design of a material whose viscosity
increases upon heating. In particular, we show how a simple design of a
binary mixture of limited valence particles can indeed provide a model
where the competition between entropy and potential energy causes the
system to show a re-entrant behavior, passing reversibly from a fluid
to a gel and again to a fluid when $T$ is varied.

It has been shown that patchy colloidal particles with a
limited number $f$ of attractive patches progressively cluster
when cooled, leading such system to form a percolating network
that at sufficiently low $T$ will incorporate all particles in the
system~\cite{emptyliquids,bianchi23,lorenzomolphys,johnjcp,kob_sastry_prl,saw:164506}.
During this process, the diffusion coefficient  progressively decreases
and the viscosity simultaneously increases. In the network state the
lifetime of the bonds (a $T$-controlled quantity) fixes the timescale
over which the system behaves as a solid. It is known that when $f=4$,
the particles (here called $A$ species) form a random tetrahedral network
which closely resembles the structure of network-forming atomic systems
like silica and silicon~\cite{binder_11}. To melt the network at low
$T$, we consider the addition of a second species ($B$) with a single
bonding patch that competes for bonding with the patches on the network
forming $A$-species.  The idea is to design a competitive mechanism
such that the bonding between $A$- and $B$-particles becomes dominant,
{\it but only} at a temperature much lower than the one at which the
$AA$-network is formed.  As a result, the stable low-$T$ phase consists
of $A$-particles decorated with $f$ $B$-particles which are free to
diffuse in the sample volume, whereas at intermediate $T$ the system
forms a highly viscous $AA$-network that is progressively fragmented
and transformed into a fluid upon heating.

\begin{figure}
\begin{center}
\begin{tabular}{ccc}
{\bf a)\hfill}\hfill & \hspace{0.1cm} & {\bf b)\hfill}\hfill\\
\centering \includegraphics[width=0.16\textwidth]{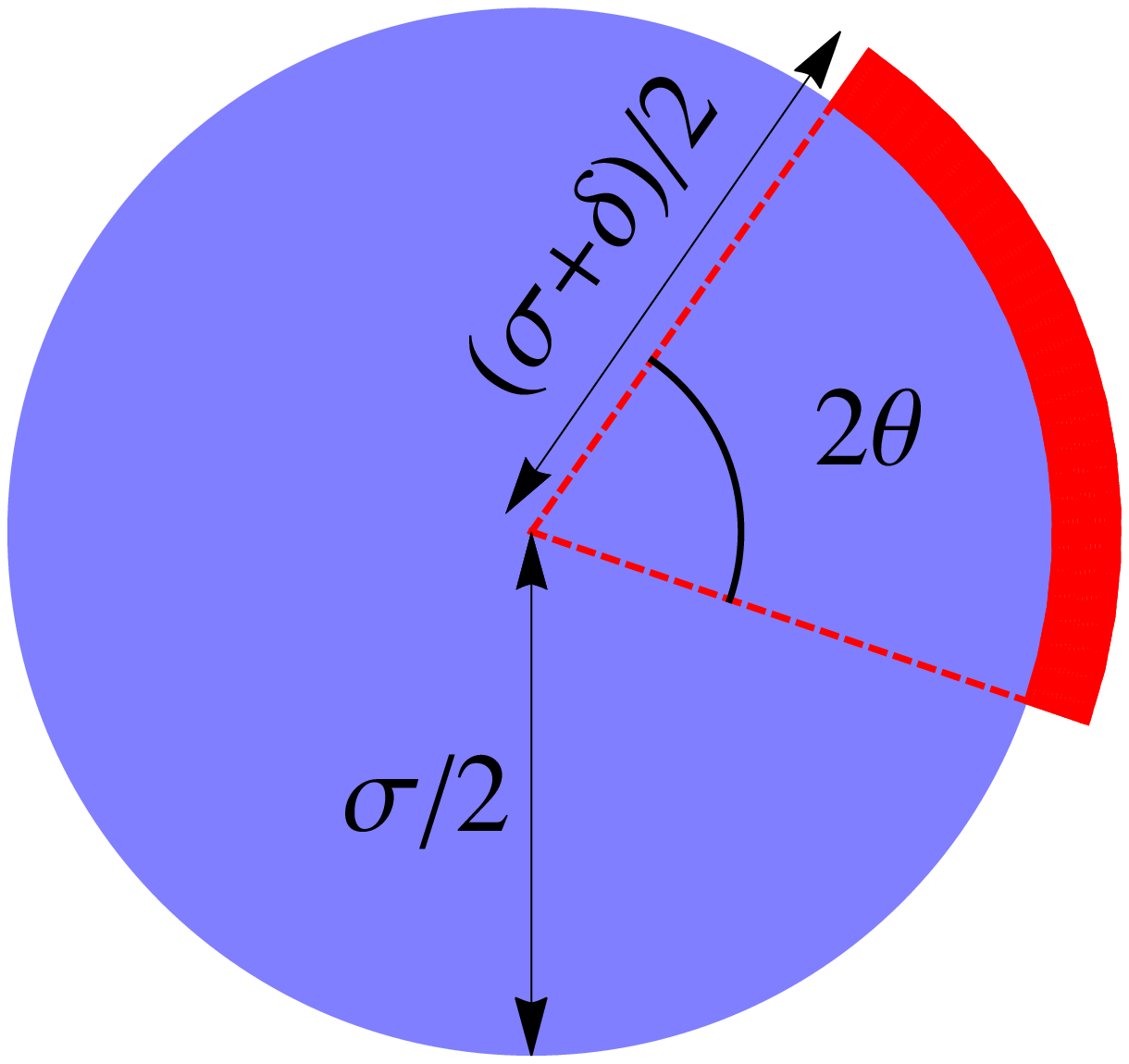} & & \includegraphics[width=0.2\textwidth]{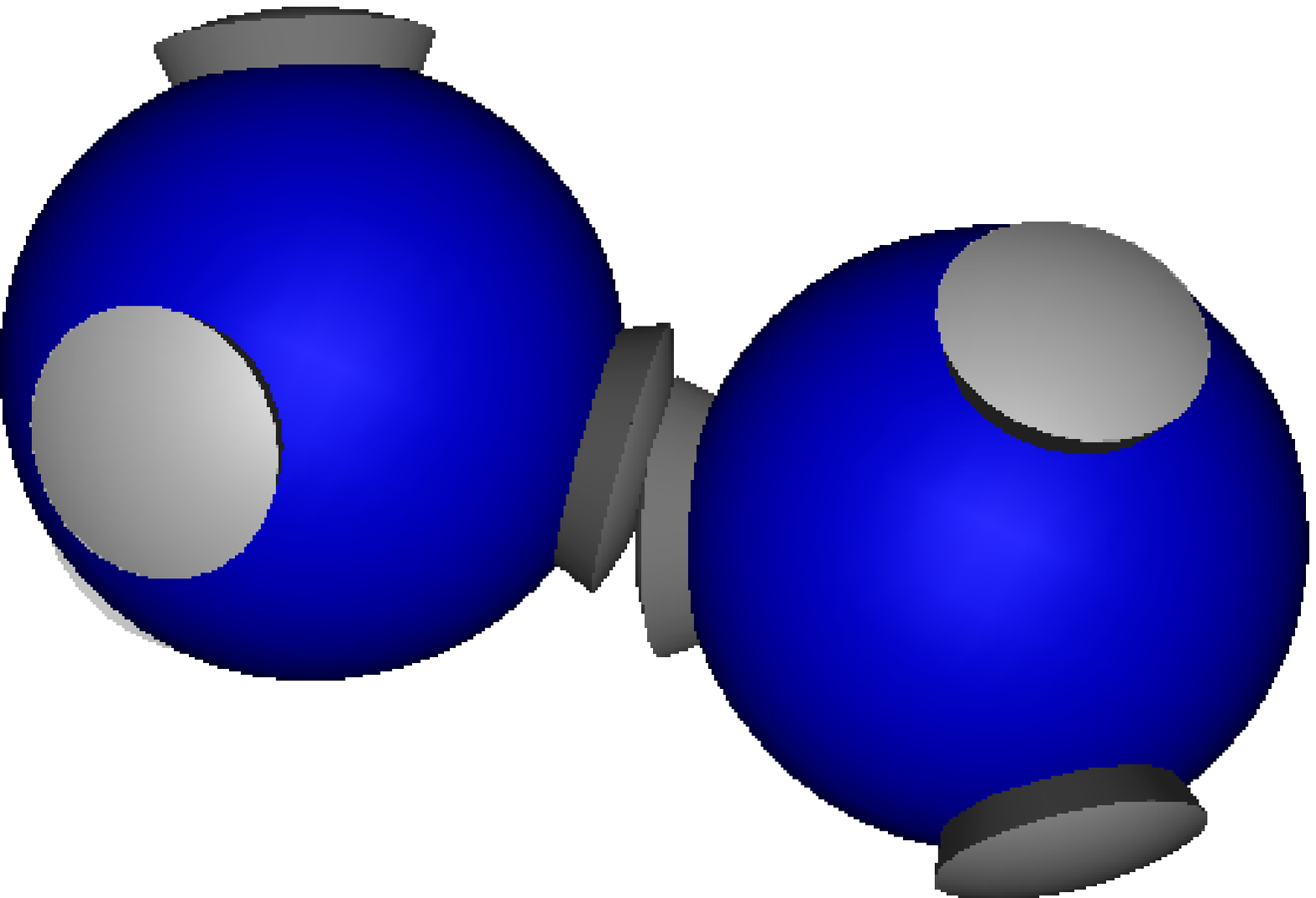} \\ 
{\bf c)\hfill}\hfill & & {\bf d)\hfill}\hfill\\
\includegraphics[width=0.14\textwidth]{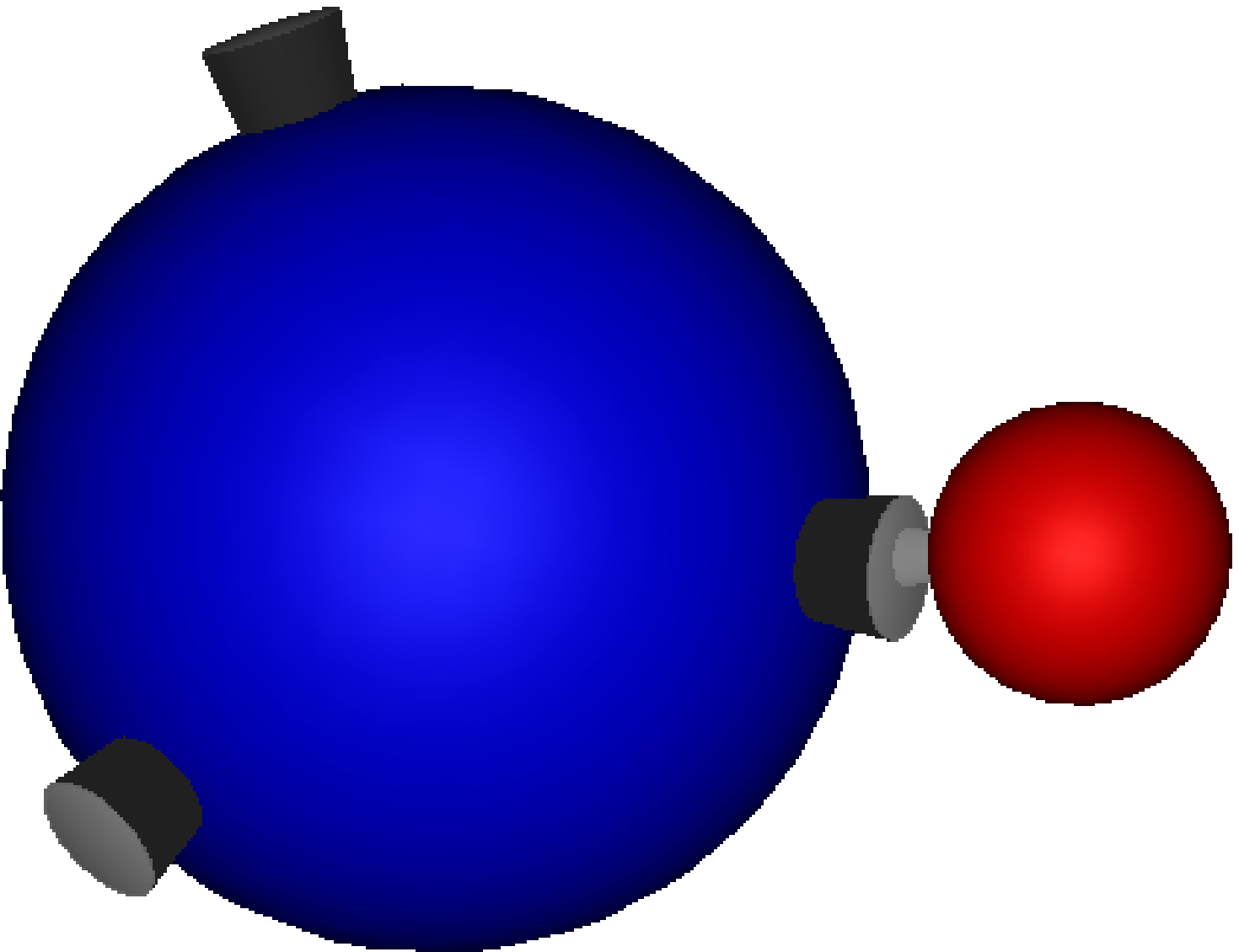} & &\centering \includegraphics[width=0.155\textwidth]{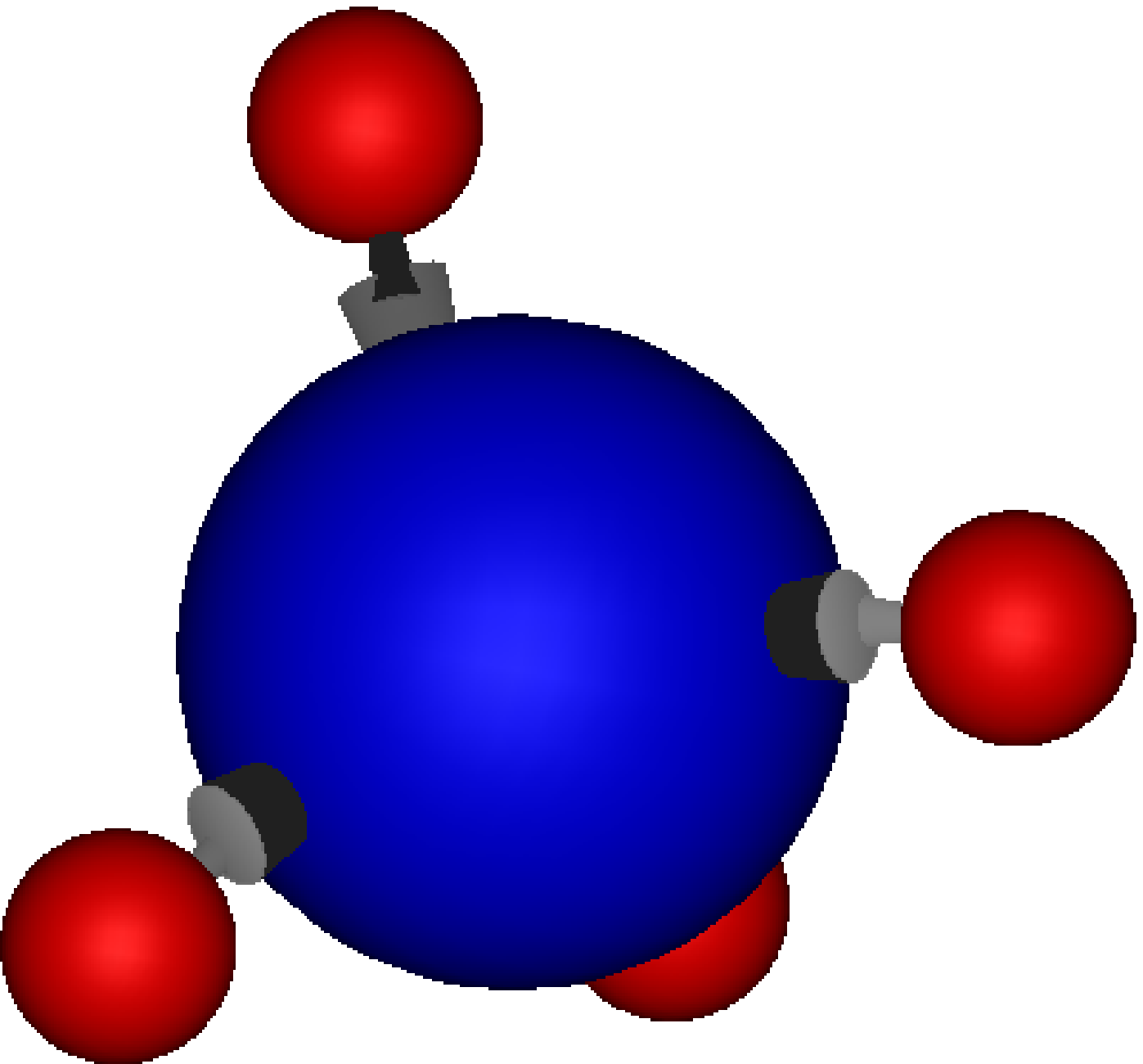} 
\end{tabular}
\caption{ a) Schematic of the interaction parameters in the Kern-Frenkel
model. An $A$-patch can bond with either another $A$-particle (b), or
with a $B$-particle (c). Panel (d) shows a ``flower'', i.e. a fully bonded
cluster consisting of one $A$-particle and four $B$-particles, representing
the lowest-energy state of the system. Here the interaction ranges and
the angular patch widths are $\cos \theta_{AA} = 0.92$, $\delta_{AA} =
0.15\sigma_A$, and $\cos \theta_{AB} = 0.99$, $\delta_{AB} = 0.2\sigma_A$.
With these choices ${\cal V}_{AA}=3.49\cdot 10^{-3} \sigma_A^{3}$ and ${\cal V}_{AB}=3.79\cdot 10^{-5} \sigma_A^{3}$ \cite{SI}. 
Note that in the binary Kern-Frenkel model, $\theta$ and $\delta$ of
each bond are defined by the species of \textit{both} bonding partners
and are not properties of individual particles.  }
\label{fig1_cartoon}
\end{center}
\end{figure}

{\it Model:} We consider a binary mixture of patchy colloids
where each $A$-particle has $f=4$ patches on its surface that are
arranged in a tetrahedral geometry and the $B$-particles have
only one patch (see Fig.~\ref{fig1_cartoon}). The patch-patch
interaction is modeled via a Kern-Frenkel potential~\cite{KF},
a model that has been extensively used over the last decade to
compare simulations and experiments on the self-assembly of patchy
colloids~\cite{granick,romanosoftmatter,achillebook}.  Each $A$-patch
can interact either with a $B$-patch with unit energy $\epsilon_{AB}$
and bonding volume ${\cal V}_{AB}$ or with another $A$-patch with
energy $\epsilon_{AA}= 0.95 \epsilon_{AB}$ and ${\cal V}_{AA}$ (${\gg
\cal V}_{AB}$). No $BB$-bonding is allowed. The bonding volumes are
each determined by an interaction range $\delta_\alpha$ and an angular
patch width $\theta_\alpha$ ($\alpha\in\lbrace$AA$,$AB$\rbrace$) (see
Fig.~\ref{fig1_cartoon} and Ref. \cite{SI}). The attractive patch-patch interaction
is complemented by an isotropic hard-core repulsion, where the spherical
cores have diameters $\sigma_{A}$ and $\sigma_{B}=0.35\sigma_A$. The size
ratio was chosen such that the $B$-particles can block the $A$-patches
from bonding to other $A$-patches without significantly increasing the
packing fraction of the pure $A$ system. Due to geometric constraints,
each patch can be involved in only a single bond.

Using event-driven molecular dynamics simulations \cite{SI,rapaport,edmdanisotropic,widepatches}, we have
studied a system of $N_A=600$ and $N_B=2400$ particles, corresponding
to a total number density $\rho \sigma_A^3=3.0$, with partial number
densities $\rho_A \sigma_A^3=0.6$ and $\rho_B \sigma_A^3=2.4$,  for
a wide range of  $T$, whose unit is given by $\epsilon_{AB}/k_B$,
where   $k_B$ is  Boltzmann's constant. The composition of the system
is thus fixed at $x_A = 0.2$. The density $\rho_A \sigma_A^3=0.6$ of
$A$-particles corresponds to the optimal density at which tetrahedral
particles form an unstrained fully bonded network\cite{simone}.
With this composition, the fully bonded network has an energy of
$2 N_A \epsilon_{AA}$, whereas a configuration in which all the
$B$-particles are bonded to the $A$-particles has a significantly
lower energy of $4 N_A \epsilon_{AB}$. We also simulate for a low $T$
($=0.04\epsilon_{AB}/k_B$) a reference system composed of 600 {\it
flowers} (see Fig.~\ref{fig1_cartoon}d), i.e.  $A$-particles bonded to
four $B$-particles. At this low $T$ no bond breaking events take place
within the simulation time.

\begin{figure}[tb]
\includegraphics[width=0.9\linewidth]{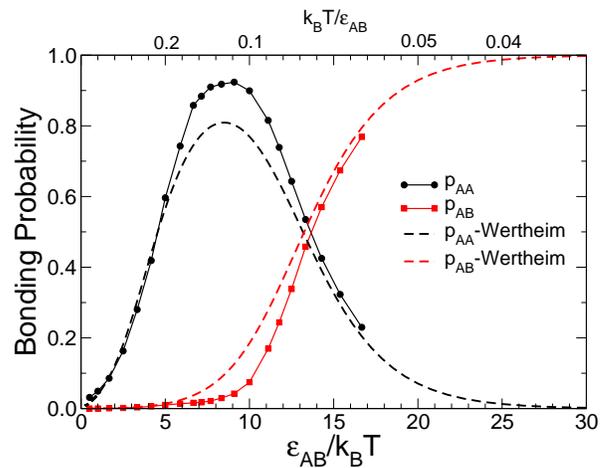}
\caption{
Probability that a patch on an $A$-particle is bonded to another
patch on an $A$-particle, $p_{\rm AA}$ (black circles), or to a patch
on a $B$-particle, $p_{\rm AB}$ (red squares). The dashed lines are the
prediction for these probabilities as obtained from the Wertheim theory.
}
\label{fig2_P_bond}
\end{figure}

{\it Results:}  Figure ~\ref{fig2_P_bond} demonstrates the basic
mechanism of the competitive interactions present in our system.
It shows the probability that a patch on an $A$-particle is bonded
to another $A$-patch ($p_{\rm AA}$) or to a $B$-patch ($p_{\rm AB}$)
as a function of the inverse $T$.  On cooling,  $p_{\rm AA}$ starts
to grow, signaling the onset of the network formation, reaching a
maximum around $T=0.11\epsilon_{AB}/k_B$.  We recall that within a
mean-field description, percolation of particles with valence $f=4$
takes place at $p_{\rm AA}=1/3$~\cite{florybook}. Since at the maximum
we find $p_{\rm AA} \approx0.9$ we can conclude that at this $T$ the
$A$-particles have formed a highly bonded percolating network. Upon
further cooling, $p_{\rm AB}$ significantly increases, showing that
the $AB$-bonds are starting to replace the $AA$ ones, i.e. while
$p_{\rm AB}$ approaches 1.0, $p_{\rm AA} \rightarrow 0$. The entropy
associated with the larger bonding volume ${\cal V}_{AA}$ for the $AA$
interaction is crucial for promoting the formation of a large number
of $AA$-bonds at intermediate $T$, before the energetically preferred
but entropically disfavored $AB$-bonds set in. Fig.~\ref{fig2_P_bond}
also shows the parameter-free theoretical predictions for the bonding
probabilities as obtained from the first-order thermodynamic perturbation
theory developed by Wertheim~\cite{wertheim1,wertheim2,delaseras_soft_Wertheim} (details on the
Wertheim calculations are reported in \cite{SI}).  The Wertheim theory
nicely captures the mechanism of competing interactions, reproducing the
position and height of the maximum of $p_{\rm AA}$ as well as the low $T$
trends of $p_{\rm AA}$ and $p_{\rm AB}$.

\begin{figure}[tb]
\includegraphics[width=0.9\linewidth]{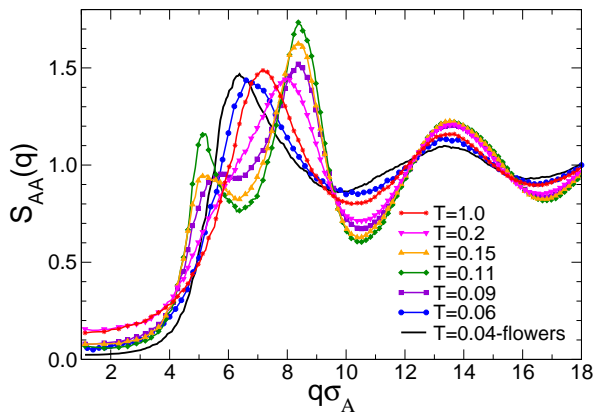}
\caption{
Partial structure factor $S_{\rm AA}(q)$ for different values of $T$
(solid lines with different symbols). The $S_{\rm AA}(q)$ for
a fluid of flowers at $T=0.04\epsilon_{AB}/k_B$ (black solid line) is also represented.
}
\label{fig3_sq}
\end{figure}

Figure \ref{fig3_sq} shows the unusual $T$ dependence of the structure
of the system, which also reflects the non-monotonic behavior of
$p_{AA}$. At high $T$, the partial structure factor $S_{AA}(q)$ shows
the conventional $q-$dependence found in simple liquids with a main peak
around $q\sigma_A=7.2$. Upon decreasing $T$ the main peak splits into two,
one located around $q\sigma_A=5.2$ and a higher one around $8.4$. This
double peak feature is typical of liquids that have a local tetrahedral
network structure, such as silicon or silica~\cite{binder_11}. The peak
at $q\sigma_A\approx 8.4$ corresponds to the nearest neighbor distance
between two bonded $A$-particles, whereas the one placed around $5.2$
is associated with the second-nearest neighbors in the tetrahedral
network. Note that this double peak structure is most pronounced at
$T\approx 0.11\epsilon_{AB}/k_B$, i.e. at the $T$ at which $p_{AA}$
has a maximum (see Fig.~\ref{fig2_P_bond}) and hence the gel is
maximally connected. When $T$ is lowered even further the double
peak structure disappears and $S_{AA}(q)$ becomes again similar to the
structure factor of a fluid composed of flowers (which is also
included in Fig.~\ref{fig3_sq} as a reference). Since the size of a
flower is larger than that of an $A$-particle, the peak position at low
$T$ is to the left of the one observed at high $T$.

We now quantify the effect of the competing interactions on the dynamics
of the system and provide evidence that the change of the structure
rich in $AB$-bonds at low $T$ to the highly bonded $AA$-network
generates a slowing down of the dynamics on heating. To do this, we
calculate the mean squared displacement (MSD) for the particles of both
species and then their corresponding diffusion coefficients $D_\alpha$
($\alpha\in\lbrace$A$,$B$\rbrace$) from the long-time behavior of the
MSD via the Einstein relation \footnote{Note that the center of mass (CM)
of a single species has a non-zero velocity (which is compensated by the
CM motion of the other species). To obtain meaningful results for the MSD
of a single species, we subtract the CM drift of the species in question
before evaluating the MSD.}. To subtract the trivial trend originated
from the $T$-dependence of the thermal velocity we divide $D_\alpha$
by a reference diffusion coefficient $D_0\equiv\sigma_A^2/\tau_0$,
where $\tau_0=\sqrt{m_A\sigma_A^2/k_BT}$ and $m_A$ is the mass of an
$A$-particle.

\begin{figure}[tb]
\includegraphics[width=0.9\linewidth]{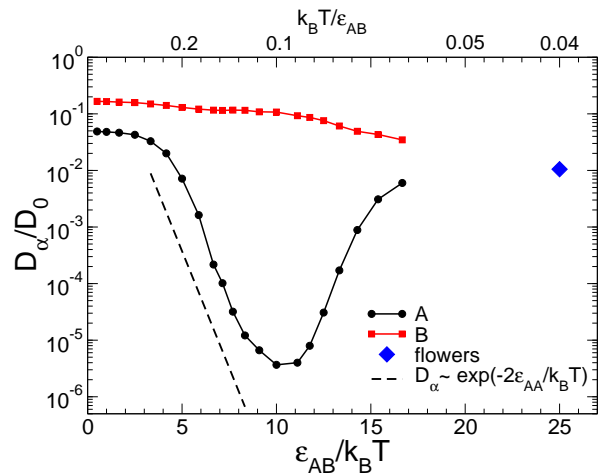} 
\caption{
Arrhenius plot of the normalized diffusion coefficient $D_\alpha /D_0$
($\alpha=A$, circles, $\alpha=B$, squares). Also included is an Arrhenius
law with activation energy $2\epsilon_{AA}$ (dashed line). The diffusion
coefficient of a fluid of flowers at low $T$ is represented by a
blue diamond.
} 
\label{fig4_diffus} 
\end{figure}

Figure \ref{fig4_diffus} shows the $T$-dependence of $D_\alpha$ in
an Arrhenius plot. At high $T$, $D_\alpha$ is approximately constant
for both type of particles, indicating that bonds do not play a
significant role. On cooling, $D_A$ starts to decrease very rapidly,
with a super-Arrhenius $T$-dependence reminiscent of that observed in
molecular networks~\cite{binder_11}, turning into an Arrhenius law
with an activation energy approximately equal to $2\epsilon_{AA}$
(see dashed line in Fig.~\ref{fig4_diffus}). Similar values of the
activation energy are typically found in tetrahedral network-forming
systems where most of the particles belong to the percolating cluster,
and bond breaking is the bottleneck for relaxation~\cite{delgado,
delgado_kob_prl, kob_sastry_prl}. Before the gel starts to decompose
at a temperature below $T\approx 0.11\epsilon_{AB}/k_B$, $D_A$ has
already decreased by four orders of magnitude compared to its value
at high $T$, indicating the formation of a persistent network. For
$T\lesssim 0.11\epsilon_{AB}/k_B$, $D_A$ starts to increase. This rising
persists down to the lowest $T$ at which we were able to equilibrate
the system. We emphasize that this non-monotonic $T$-dependence is only
observed for the $A$-particles, i.e. the particles which are involved in
the formation of the network. In contrast,  $D_B$  shows only a rather
mild $T$-dependence. Figure \ref{fig4_diffus} also shows the diffusion
coefficient of the fluid of flowers at $T=0.04\epsilon_{AB}/k_B$ for
which $D_A=D_B$.  This common value of the diffusion coefficient provides
a low-$T$ limit to which both $D_A$ and $D_B$  converge, consistent with
the trends shown by $D_A$ and $D_B$ at low $T$.

\begin{figure}[tb]
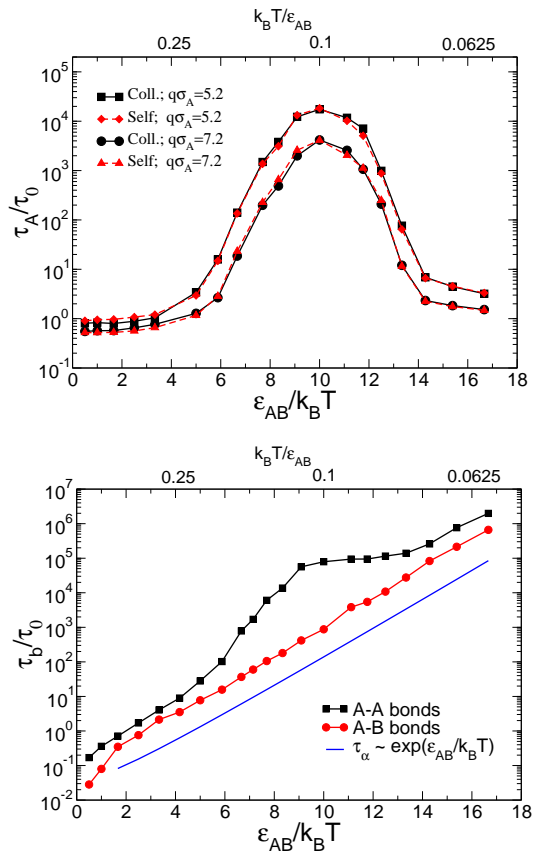

\includegraphics[width=0.8\linewidth]{fig5a_relaxtime.eps} \vspace{0.3cm}

\includegraphics[width=0.8\linewidth]{fig5b_relaxtime.eps}
\caption{
a) Arrhenius plot of the normalized relaxation time $\tau_A(q)/\tau_0$
as obtained from the self (dashed lines) and collective (solid lines)
scattering functions, where $\tau_0=\sqrt{m_A\sigma_A^2/k_BT}$. The
different curves correspond to the wave-vectors given by the first two
peaks in $S_{AA}(q)$. b) Arrhenius plot of the normalized bond-persistence
time $\tau_b/\tau_0$ for the AA (squares) and AB bonds (circles). Also
included is an Arrhenius law with activation energy $\epsilon_{\rm AB}$
(blue solid line).
}
\label{fig5_relaxtime}
\end{figure}

A non-monotonic behavior of the characteristic time is also found in
the time evolution of the  collective- and self-intermediate scattering
functions. Their study provides insight into how the relaxation dynamics
depends on the considered length scale.  Figure ~\ref{fig5_relaxtime}a
shows an Arrhenius plot of the  relaxation time $\tau_A(q)$ determined
from the time integral of the intermediate scattering function of
the $A$-particles for two different $q-$vectors: $q\sigma_A=5.2$
and $q\sigma_A=7.2$, which correspond, respectively, to the location
of the first peak in the network and in the high-$T$ fluid (see
Fig.~\ref{fig3_sq}). We find that the self and collective relaxation
times $\tau_A(q)$, normalized by $\tau_0$, show qualitatively the
same $T$-dependence: a plateau at high $T$, a fast increase within
the $T$-range in which the network is formed, a quick decrease
once the network starts to break up again, and a final plateau at
low $T$. This $T$-dependence is observed for both values of $q$,
indicating that the relaxation mechanism does not depend  on the length
scale considered. Analogous to the diffusion coefficient, the self and
collective dynamics of the $B$-particles are found to be faster than
those of the $A$-particles and their $q$ and $T$ dependence will be
reported elsewhere.

To provide further evidence that the system is ergodic on long time
scales, i.e. that the structure of the system has completely lost
its memory of the initial state, we investigate the bond persistence
function $p_b(t)$, i.e. the probability that a bond which is present at
time zero is also present at time $t$.  When $p_b(t)$ approaches zero,
all bonds which were present at time zero have been broken. Hence, the
relaxation time of $p_b(t)$ provides information on the restructuring
time of the network connectivity. Figure \ref{fig5_relaxtime}b shows
an Arrhenius plot with the $T$-dependence of the  decay time $\tau_b$,
where  $p_b(\tau_b)=e^{-1}$. At intermediate and low $T$, $\tau_b$ is
larger than the relaxation times shown in Fig.~\ref{fig5_relaxtime}a.
We find thus that $p_b(t)$  decays to zero only on a time scale that is
significantly longer than the relaxation times associated with
the scattering functions, confirming that some fraction of spacial
decorrelation of the network, as quantified by the collective scattering
function, takes place at partially fixed bonding pattern. In other words,
while the decay to zero of the scattering functions for a given $q-$value
implies that the particles have moved over a distance on the order of
$2\pi/q$, such motion does, however, not necessarily require that {\it
all} bonds are broken, since, e.g., a cluster of particles can move
in a collective manner. A bond can persist up to very long times even
if the structure of the system changes significantly.  The extreme
case  occurs at very low $T$ where the system is a fluid of flowers
that relaxes relatively quickly but in which  $AB$-bonds survive for a
very long time. Figure \ref{fig5_relaxtime}b also shows the different
$T$-dependence of the lifetime of the $AA$- and of the $AB$-bonds.
An Arrhenius dependence with an activation energy very close to
$\epsilon_{\rm AB}$ (see dashed line in Fig.~\ref{fig5_relaxtime}b)
characterizes the $AB$-bonds in the entire $T$-range, suggesting that
the mechanism for the breaking of an $AB$-bond is not collective in
nature but a simple activated process. In contrast to this behavior,
the breaking time for an $AA$-bond follows an Arrhenius law at high $T$
but  becomes super-Arrhenius within the $T$-region in which the gel
forms, showing that the bond-breaking process becomes coupled to the
degree of bonding.  At low $T$, when most of the $A$-sites are bonded
to $B$-particles and the network is disrupted,  the effective activation
energy decreases again to recover at very low $T$ an Arrhenius behavior
with an activation energy given by $\epsilon_{\rm AA}$.

{\it Conclusions:} In summary, we have shown that a judicious  choice of
the interaction parameters of a binary  mixture of $A$ and $B$ patchy
particles allows us to generate a non-monotonic  $T$-dependence of
its dynamic properties.  Essentially, we set up a competition between
network-forming $AA$-bonds and network-breaking $AB$-bonds, and tune
the balance between both bond types by choosing their bonding volumes
(and therefore the entropy) and energies. At high $T$, very few bonds
are formed, and the system behaves similar to a binary fluid of hard
spheres. At slightly lower $T$, the stability of a stiff, percolating
network is ensured by the larger entropy associated with the much larger
bonding volume $\cal V_{AA}$. However, when $T$ is decreased even further,
the system instead forms a fluid of small clusters, stabilized by a much
larger number of $AB$-bonds, corresponding to a lower potential energy. We
have shown that, compared to both the high-$T$ and low-$T$ fluids,
the network state relaxes and diffuses more slowly by several orders
of magnitude. Thus, the system forms a reversible gel that transforms into a fluid
upon both heating and cooling.  We conclude by listing  two experimental
systems which we believe are very promising candidates to experimentally
test the ideas presented in this Letter: a solution of DNA constructs
of valence four\cite{dnatetramersJACS,dnabellini} in the  presence of
competing DNA single strands and the binary mixture of patchy particles
recently synthesized\cite{dnapatchy1}.  Both these model systems have
the potential to provide soft materials that gel on heating.

{\it Acknowledgements:} 
We acknowledge funding received from COMPLOIDS and ERC Grant agreement No. 226207.
W. Kob acknowledges support from the Institut Universitaire de France.

\bibliographystyle{apsrev}

\end{document}